\documentstyle[proceedings,epsf]{crckapb} 

\newcommand{\HI}{H\thinspace\protect\footnotesize I\protect\normalsize} 
\newcommand{\HIb}{H\thinspace\protect\footnotesize\bf I\protect\normalsize}
\newcommand{\hi}{H\thinspace{\protect\scriptsize I}}

\newcommand{\ij}{\mbox{$I-J$}}
\newcommand{\jk}{\mbox{$J-K$}}
\newcommand{\B}{{$B$}}

\newcommand{\II}{{$I_c$}}
\newcommand{\J}{{$J$}}
\newcommand{\K}{{$K_s$}}
\newcommand{\tfr}{Tully\,--\,Fisher relation}

\newcommand{\kms}{\,km\,s$^{-1}$}
\newcommand{\etal}{{\it et~al.}}
\newcommand{\cf}{{\it cf.\,}}
\newcommand{\eg}{{\it e.g.},\ }         
\newcommand{\ie}{{\it i.e.},\ }         
\newcommand{\ca}{\citeauthor}
\newcommand{\cy}{\citeyear}
\newcommand{\shc}{\shortcite}

\def\la{\mathrel{\hbox{\rlap{\hbox{\lower4pt\hbox{$\sim$}}}\hbox{$<$}}}}
\def\ga{\mathrel{\hbox{\rlap{\hbox{\lower4pt\hbox{$\sim$}}}\hbox{$>$}}}}

\def\deg{{^\circ}}
\def\arcmin{\hbox{$^\prime$}}

\def\fm{\hbox{$.\!\!^{\rm m}$}}

\def\fdg{\hbox{$.\!\!^\circ$}}
\def\farcm{\hbox{$.\mkern-4mu^\prime$}}

\def \aa#1#2   {{\em Astr. Astrophys. \/} {\bf #1}, {#2}}
\def \aas#1#2  {{\em Astr. Astrophys. Suppl. Ser. \/} {\bf #1}, {#2}}
\def \aj#1#2   {{\em Astron. J. \/} {\bf #1}, {#2}}
\def \apj#1#2  {{\em Astrophys. J. \/} {\bf #1}, {#2}}
\def \apjs#1#2 {{\em Astrophys. J. Suppl. Ser. \/} {\bf #1}, {#2}}
\def \mnras#1#2{{\em MNRAS \/} {\bf #1}, {#2}}
\def \nat#1#2  {{\em Nature \/} {\bf #1}, {#2}}

\begin{opening}
\title{Large-Scale Structures Behind the Milky Way from Near-IR
 Surveys}

\author{R.C. Kraan-Korteweg}
\institute{DAEC, Observatoire de Paris, Meudon, France, and\\
           Dept. de Astronomia, Universidad de Guanajuato, Mexico}
\author{A. Schr\"oder}
\institute{Institute of Astronomy, NCU, Chung-Li, Taiwan}
\author{G.A. Mamon}
\institute{IAP, Paris, France, and \\
           DAEC, Observatoire de Paris, Meudon, France}
\author{S. Ruphy}
\institute{DESPA, Observatoire de Paris, Meudon, France}

\end{opening}

\begin{document}

\begin{abstract}
About 25\% of the optical extragalactic sky is obscured by the dust and stars
of 
our Milky Way. Dynamically important structures might still lie hidden in this
zone. Various approaches are presently being employed to uncover the galaxy
distribution in the Zone of Avoidance (ZOA) but all suffer from (different)
limitations and selection effects.

We investigated the potential of using the DENIS NIR survey for studies of
galaxies behind the obscuration layer of our Milky Way and for mapping the
Galactic extinction. As a pilot study, we recovered DENIS \II, \J\
and \K\ band images of heavily obscured but optically still visible galaxies.
We determined the \II, \J\ and \K\ band luminosity functions
of galaxies on three DENIS strips that cross the center of the 
nearby, low-latitude, rich cluster Abell 3627. The 
extinction-corrected \ij\ and \jk\ colours of these
cluster galaxies compare well with that of an unobscured cluster.
We searched for and identified galaxies at latitudes where the Milky Way 
remains fully opaque ($|b| < 5\deg$ and $A_B \ga 4-5^{\rm m}$) 
--- in a systematic search as well as around positions of galaxies detected
with the  
blind \HI -survey of the ZOA currently
conducted with the Multibeam Receiver of the Parkes Radiotelescope.
\end{abstract}

\section{Introduction }

Some of the results of this study have already been reported in \ca{denis1}
\cy{denis1} (Paper I). For a comprehensive
description, the goals and earlier results of this project are
repeated here, but the reader is referred to paper I for
details on earlier presented results.

About 25\% of the optically visible extragalactic sky is obscured by the dust
and stars of our Milky Way. Dynamically important structures --- individual
nearby galaxies (\cf\ \ca{Dw1} \cy{Dw1}) as well as large clusters
and superclusters (\cf\ \ca{A3627} \cy{A3627}) --- might still lie
hidden in this zone.
Complete whole-sky mapping of the galaxy and mass distribution is
required in explaining the origin of the peculiar velocity of the
Local Group and the dipole in the Cosmic Microwave
Background.

Various approaches are presently being employed to uncover the galaxy
distribution in the ZOA: deep optical searches, far-infrared
(FIR) surveys (\eg IRAS), and blind \HI\ searches. All methods produce new
results, but all suffer from (different) limitations and selection
effects. Here, the near infrared (NIR) surveys such as 2MASS \cite{2m}
and DENIS in the southern sky, \cite{den,denmes} could provide
important complementary data. NIR surveys will:\\
\indent $\bullet$ be sensitive to early-type galaxies --- tracers of
massive groups and clusters --- which are missed in IRAS and \HI\ surveys,\\
\indent $\bullet$ have less confusion with Galactic objects compared to FIR
surveys,\\
\indent $\bullet$ be less affected by absorption than optical surveys.\\
But can we detect galaxies and obtain accurate magnitudes in crowded
regions and at high foreground extinction using NIR surveys? To assess
the performance of the DENIS survey at low Galactic latitudes we
addressed the following questions:

(1) How many galaxies visible in the $B_J$ band ($B_{\rm lim} \approx 19\fm0$)
  can we recover in \II\ ($0.8\mu \rm m$), \J ($1.25\mu \rm m$) and \K
  ($2.15\mu \rm m$)? Although less affected by extinction (45\%, 21\% and 9\%
  as 
  compared to $B_J$), their respective completeness limits are lower ($16\fm0,
  14\fm5$, and $12\fm2$, \ca{gam3} \cy{gam3,gam4}).

(2) Can we determine the \II, \J, and \K\ band
 luminosity functions?

(3) Can we map the Galactic extinction from NIR colours of galaxies
  behind the Milky Way?

(4) Can we identify galaxies at high extinction ($A_B > 4-5^{\rm
  m}$) where optical surveys fail and FIR surveys are plagued by confusion?

(5) Can we recover heavily obscured spiral galaxies detected in a
    blind \HI\ search and hence extend the peculiar velocity field
    into the ZOA via the NIR \tfr ?

We pursued these questions by comparing available DENIS data with results
from a deep optical survey in the southern ZOA (Kraan-Korteweg \& Woudt 1994,
Kraan-Korteweg \etal\ 1995, 1996, and references therein).  In this region
(\mbox{$265\deg \la \ell \la 340\deg$,} $|b| \la 10\deg$), over 11\,000
previously unknown galaxies above a diameter limit of $D\!=\!0\farcm2$ and
with $B \la 19\fm0-19\fm5$ have been identified (\cf\ Fig.~1 in Paper I).
Many of the faint low-latitude galaxies are intrinsically bright
galaxies. Within the survey region, we investigated DENIS data at what seems
to be the core of the Great Attractor (GA), \ie\ in the low-latitude
($\ell\!=\!325\deg$, $b\!=\!-7\deg$), rich cluster Abell 
3627, where the Galactic
extinction is well determined \cite{Sey}, and in its extension across the
Galactic Plane where the Milky Way is fully opaque.

\section {Expectation from DENIS}

\begin{figure} [hbt]
\centerline {\epsfxsize=9.5cm \epsfbox[20 161 564 532]{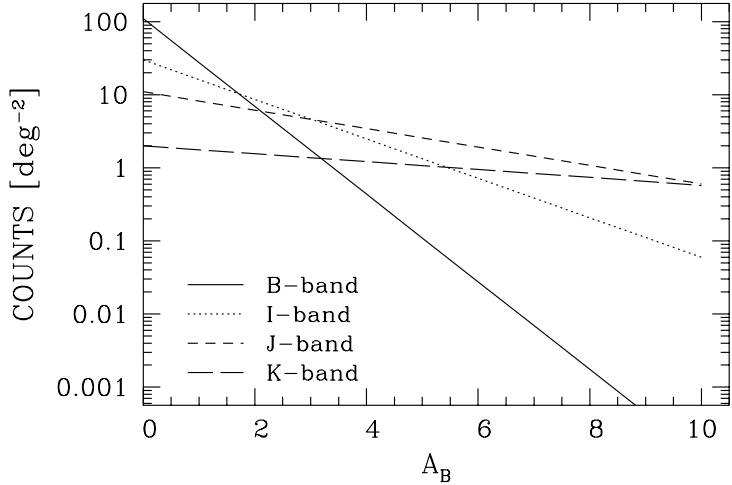}} 
\caption{Predicted galaxy counts in \B , \II , \J\ and \K\ as a function of
absorption in \B , for highly complete and reliable DENIS galaxy samples and
a $B_J \leq 
19^{\rm m}$ optical sample. }
\label{galctsplot}
\end{figure}

What are the predictions for DENIS at low latitudes? In unobscured
regions, the density of galaxies per square degree is 110 in the blue
for $B_J\le19\fm0$ \cite{gar}, and 30, 11, and 2 in the \II , \J\ and
\K\ bands for their respective completeness limits of
$I_{\rm lim}\!=\!16\fm0$, $J_{\rm lim}\!=\!14\fm0$, $K_{\rm lim}\!=\!12\fm2$
(\ca{gam3} \cy{gam3,gam4}).
The number counts in the blue decrease with increasing obscuration as
$N(<\!B) \simeq 110 \times {\rm dex} (0.6\,[B-19])\,$deg$^{-2}$. According to
\ca{Car} \shc{Car}, the extinction in the NIR passbands are
$A_{I_c}\!=\!0\fm45$, $A_J\!=\!0\fm21$, and $A_{K_s}\!=\!0\fm09$ for
$A_B\!=\!1\fm0$, 
hence the decrease in number counts as a function of extinction is
considerably slower. Figure~\ref{galctsplot} shows the predicted surface
number density  of
galaxies for DENIS and for $B < 19$, as a
function of Galactic foreground extinction.

The NIR becomes notably more efficient at $A_B\simeq 2-3^{\rm m}$, while
the Milky Way becomes opaque at $A_B \ge 4^{\rm m}$. At an extinction of $A_B
\simeq 6^{\rm m}$, \J~and \K\ become superior to the \II\ band, and we can
expect to 
find galaxies in \J\ and \K, even at $A_B \!=\! 10^{\rm m}$. These are very
rough predictions and do not take into account any dependence on morphological
type, surface brightness, orientation and crowding, which will surely lower the
counts of actually detectable galaxies counts \cite{gam}.

In April 1997, a new cooling system for the focal instrument of DENIS has been
mounted. This appears to increase the \K\ band limiting magnitude by $\sim$ 0.5
magnitude and therewith the
number of galaxies detectable in the deepest obscuration layer of the Milky
Way by a factor of about 2. 
Consequently, 
the {\it long dashed curve\/} representing
the \K\ counts in Figure~\ref{galctsplot} should be moved up by roughly a factor
of 2, which
would make the \K\ passband competitive with \J\ starting at $A_B \simeq
7^{\rm m}$. 

\section{DENIS-data in the Norma cluster A3627}
\subsection{Recovery of galaxies found in the \B\/ band } \label{recov}

Three high-quality DENIS strips cross the cluster Abell 3627
practically through its center. We inspected 66 images
which cover about one-eighth of the cluster area within
its Abell-radius of $R_A = 1\fdg75$ (each DENIS image is
$12\arcmin$x$12\arcmin$, offset by $10\arcmin$ in declination and
right ascension). The extinction over the regarded cluster area varies
as $1\fm2 \le$ A$_B \le 2\fm0$.

We cross-identified the galaxies found in the optical survey with the
DENIS \II, \J, and \K\ images. An example of a DENIS image in
the central part of the cluster is given in Figure~3 of Paper I.
On the 66 images, 151 galaxies had been identified in the optical. We
have recovered 122 galaxies in the \II\ band, 100 in the \J\ band, and
74 in the \K\ band (not including galaxies visible on more than one
image). As suggested by Figure~\ref{galctsplot}, the \K\
band indeed is not optimal for identifying obscured galaxies at
these latitudes due to its shallow magnitude limit.  Most
of the galaxies not re-discovered in \K\ are low surface brightness
spiral galaxies. 

Surprisingly, the \J\ band provides better galaxy detection than the \II\ band.
In the latter, the severe star crowding makes identification of faint
galaxies very difficult. At these extinction levels, the optical survey does
remain the most efficient in {\it identifying} obscured galaxies.

\subsection{Photometry of galaxies in the Norma cluster }

We have used a preliminary galaxy pipeline \cite{gam3,gam4}, based upon the
SExtractor package \cite{bertinarnouts}
on the DENIS data in the Norma cluster to obtain \II , \J\
and \K\ Kron photometry.  Although many of the galaxies have a considerable
number of stars superimposed on their images, magnitudes derived
from this fairly automated algorithm agree well with the few known, independent
measurements.

\begin{figure} [ht]
\vspace{-4.5cm}
\centerline {\epsfxsize=12.cm \epsfbox{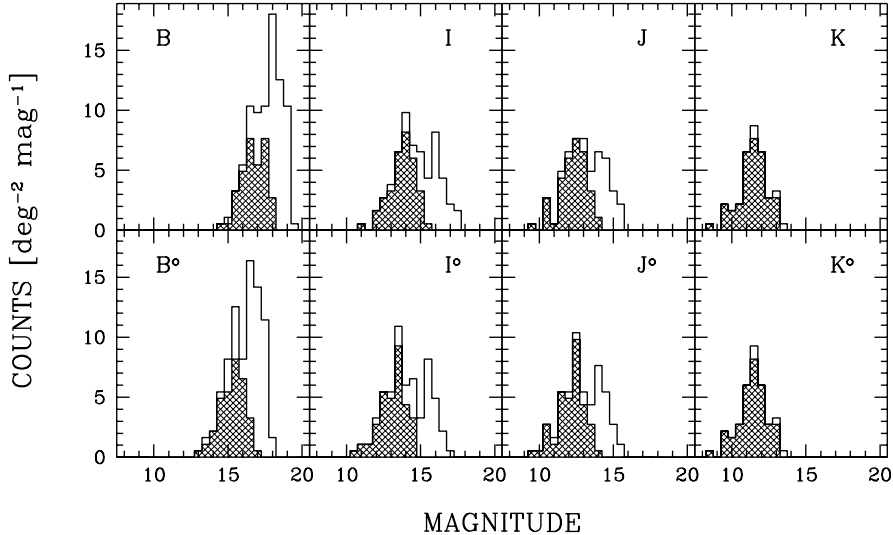}}
\caption{The luminosity function for the observed Norma galaxies in \B, \II,
\J, 
and \K. The bottom panels display magnitudes corrected for foreground
extinction. The {\it hashed histograms\/} represent the sample common to all 4
passbands ($N=60$).}
\label{lfnorplot}
\end{figure}

Magnitudes could be determined for 109, 98 and 64 galaxies of the 122,
100, 74 galaxies re-discovered in \II, \J, and \K. Figure~\ref{lfnorplot} shows
the luminosity function (LF) of these galaxies together with the \B\ band
LF of the 151 galaxies visible on the same 66 DENIS
images. The histograms are normalised to the area covered
by the 66 images. The hashed area marks the 60 galaxies common to all
4 passbands. This subsample is mainly restricted by the \K\ band.
The magnitudes in the bottom row are corrected for extinction. The
corrections are derived from Mg$_2$-indices
of elliptical galaxies in the cluster (Woudt \etal\ in prep.)
and interpolations according to the Galactic \HI\ distribution.

To assess whether the LFs displayed here are, in fact, representative of the
cluster as a whole --- and therefore the extinction corrected NIR \II, \J,
and \K\ band LFs displayed in the lower panels characteristic for rich
clusters --- we compared the \B\ band LF of the 151 galaxies on the 66
DENIS-images with the cluster LF as a whole (\cf\ \ca{P_dis}, \cy{P_dis}).
The extinction-corrected blue cluster LF of the 609 galaxies within the Abell
radius, scaled to the Abell area, actually has lower number counts than the
\B$^o$ band LF displayed in the bottom panel of Figure~\ref{lfnorplot}. This is
explained by the fact that our three strips cross the center of the cluster
and therewith the region of highest density. The comparison indicates that we
are fairly complete to a magnitude of $B^o = 16\fm5$, which is more or less
the shaded area, and that the shape of the total LF is very similar to the
distribution of the common subsample.

Even though these LFs are still preliminary (we have so
far covered only a small area of the Norma cluster and will have missed
dwarf galaxies and other LSB galaxies due to the foreground
obscuration) the here determined extinction-corrected
LFs of the galaxies common to all passbands can be regarded as a first
indication of the bright end of the NIR \mbox{\II,} \J, and \K\  band
LFs in rich clusters.


{}From the below discussed colours of the Norma galaxies, we know that
the extinction corrections are of the correct order.
Adopting a distance to A3627 of 93 Mpc \cite{A3627},
thus $m\!-\!M = 34\fm8$, the 60 galaxies cover a luminosity range 
in \K\ of \mbox{$-25\fm3 < M_K^o < -21\fm8$.} This
compares well with the bright end of the \mbox{\K\ band} LF of the Coma
cluster core derived by \ca{mobasher} \shc{mobasher}, although it remains
puzzling why the number counts derived by them (\cf\ their Table~1) 
are so much lower compared to the A3627 cluster.

The NIR magnitudes have been used to study the colour\,--\,colour diagram
\ij\ versus \jk. This has been presented and discussed in detail in Paper~I.
Here it suffices to state that the extinction-corrected colours of the
cluster galaxies match the colours of galaxies in unobscured high latitude
regions \cite{gam3} extremely well, suggesting that our preliminary
photometry is reasonably accurate.  Moreover, the shift in colour can be
fully explained by the foreground extinction or, more interestingly, the NIR
colours of obscured galaxies provide, in principle, an independent way of
mapping the extinction in the ZOA (see also \ca{gam2}, \cy{gam2}).

\section{`Blind' search for galaxies}

\begin{figure} [ht]
\hfil{\epsfxsize=12.5cm \epsfbox{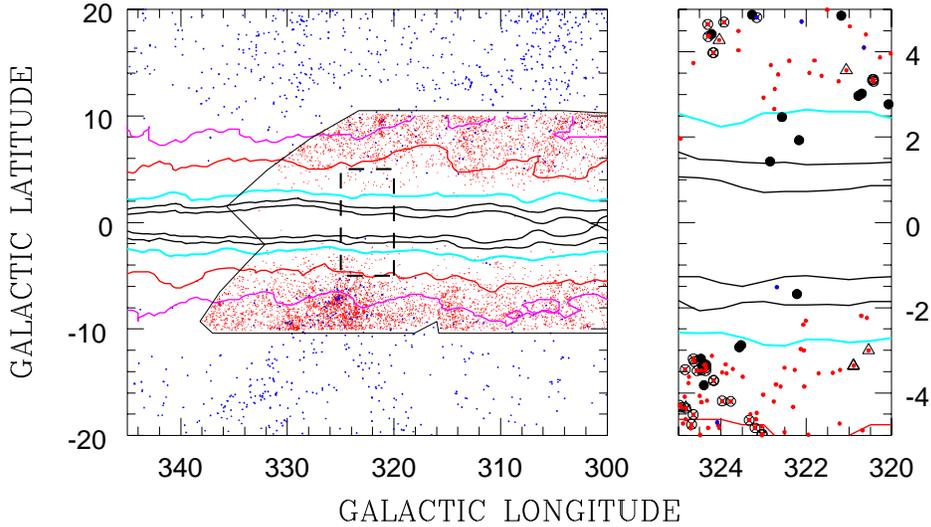}}\hfil  
\caption{Galaxy distribution in the GA region displaying Lauberts galaxies
($D \ge 1\farcm0$, Lauberts 1982) and galaxies from the deep optical search
($D \ge 0\farcm2$, outlined area). The superimposed {\it contours\/} represent
absorption levels of $A_B=1\fm5, 2\fm5, 5\fm0$ ({\it thick line\/}), $7\fm5$
and $10\fm0$, as determined from \hi\ column densities and assuming a constant
gas/dust ratio. The {\it box\/} marks the DENIS blind search area with the
results shown enlarged in the right panel: optical galaxies re-identified on
DENIS images ($N\!=\!31$, including 3 uncertain identifications) as {\it
large encircled crosses\/}, optical galaxies not seen by DENIS ($N\!=\!6$) as
{\it triangles\/}, and newly identified, optically invisible galaxies
($N\!=\!15$) as {\it filled dots\/}.}
\label{bsrchplot}
\end{figure}

The GA is suspected to cross the Galactic Plane from the Norma cluster in the
south towards the Centaurus cluster in the north. In this region, we
performed a search for highly obscured galaxies on the so far existing
DENIS survey images. The search area within the GA-region
--- marked as a {\it dashed box\/} in Figure~\ref{bsrchplot} --- is defined as
$320\deg \leq \ell \leq 325^\deg$ and $|b|\leq 5\deg$.

Of the 1800 images in this area we have inspected 385 by eye (308 in \K).
37 galaxies at higher latitudes were known from the optical survey.
28 of these could be re-identified in the \II\ band, 26 in the
\J\  band, and 14 in the \K\  band. They are plotted as {\it encircled
crosses\/} in Figure~\ref{bsrchplot}. In addition, we found 15 new galaxies
in \II\ and \J, 11 of which also appear in the \K\  band ({\it filled
circles\/}).  The ratios of galaxies found in \II\ compared
to \B, and of \K\ compared to \II\, are higher than in the Norma
cluster. This is due to the higher obscuration level (starting 
with A$_B \simeq 2\fm3 -3\fm1$ at the high-latitude border of the 
search area, \cf\ {\it contours\/} of Fig.~\ref{bsrchplot}).

On average, we have found about 3.5 galaxies per square degree in the
\II\ band. This roughly agrees with the predictions of
Figure~\ref{galctsplot}, although the number of the inspected images and
detected galaxies are too low to allow a statistical conclusion. Since we
looked in an overdense region we expect {\it a priori} more
galaxies. On the other hand, we do not expect to find galaxies below
latitudes of $b \simeq 1\deg-2\deg$ in this longitude
range \cite{gam}. The visual impression of the low-latitude
images substantiates this --- the images are nearly fully covered with stars.

\begin{figure} [htb]
\centerline {\epsfxsize=12.cm \epsfbox{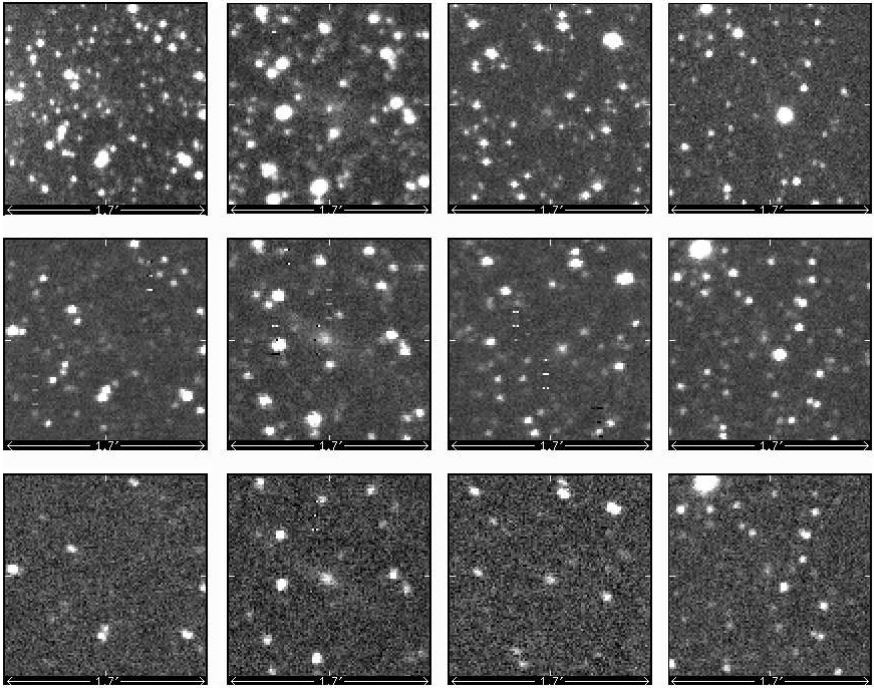}}
\caption{DENIS survey images (before bad pixel filtering) 
of four galaxies found in
the deepest extinction layer of the Milky Way; the \II\  band image
is at the {\it top\/}, \J\ in the {\it middle\/} and \K\ at the {\it
bottom\/}.} 
\label{bsexplot}
\end{figure}

Figure~\ref{bsexplot} shows a few characteristic examples of highly
obscured galaxies found in the DENIS blind search. \II\ band images are
at the top, \J\ in the middle and \K\ at the
bottom. The left-most galaxy is located at $(l,b) = (324\fdg6,-4\fdg5$),
with $A_B = 2\fm8$ as estimated from \mbox{\HI -column} densities 
\cite{kerr} following the precepts of \ca{BH} \shc{BH}. 
It is barely visible in the \J\  band, although its
\B\  band image is similar to the \B\ of the second galaxy. This galaxy
at $(l,b) = (324\fdg7,-3\fdg5$) is, however, subject to heavier extinction 
($A_B = 3\fm7$) and hence easier to recognise in the NIR. The most
distinct image is the \J\  band. The third galaxy
at even higher extinction $(l,b,A_B) = (320\fdg1,+2\fdg5,4\fm6$) is
not visible anymore in the \B\  band. Neither is the fourth galaxy:
at $b=+1\fdg9$ and $A_B = 6\fm3$ this galaxy is not even
visible in the \II\  band and very faint in \J\ and \K.

The most important result from this search is that {\it
highly obscured, optically
invisible galaxies can indeed be unveiled in the NIR\/} and --- as
indicated with the distribution in the right panel of
Figure~\ref{bsrchplot} --- found at lower latitudes than the deep optical
survey. The lowest Galactic latitude at which we found a galaxy is
$b \simeq 1.5\deg$ and $A_B \simeq 7\fm5$.

\section{Galaxies detected in \HIb }
 
NIR surveys are the only tools that will identify 
early-type galaxies and therewith uncover the 
cores of massive groups and clusters at very low-latitudes.
In addition, highly obscured spiral galaxies should be detectable with 
these surveys as well. Such identifications will proof important
in connection with the systematic blind \HI\ survey currently
conducted with the Multibeam Receiver (13 beams in the focal plane
array) at the 64\,m Parkes telescope: a deep survey with a $5 \sigma$
detection limit of 10\,mJy is being performed in the most opaque
region of the southern Milky Way ($213\deg \la \ell \la
33\deg$; $|b| \la 5\deg$) for the velocity range of
$-1000 \la v \la $12000 \kms \cite{MB}. Roughly
3000 detections are predicted. Hardly any of 
them will have an optical counterpart. However, at these latitudes 
many might be visible 
in the NIR. The combination of data from these two surveys, \ie\ 
NIR photometry with \HI -data (velocity and linewidth) will
proof particularly interesting because it will allow 
the extension of peculiar velocity data {\it into} the ZOA
via the NIR \tfr.

Only a few cross-identifications were possible with the data
available from both surveys by June 1997. But we could identify thirteen
galaxies detected blindly in \HI\ on existing DENIS images. Four of 
them are visible in the \B, \II, \J, and \K\  bands. The other 
galaxies are only seen in the NIR. Four of them need 
further confirmation.

\begin{figure} [htb]
\centerline {\epsfxsize=12.cm \epsfbox{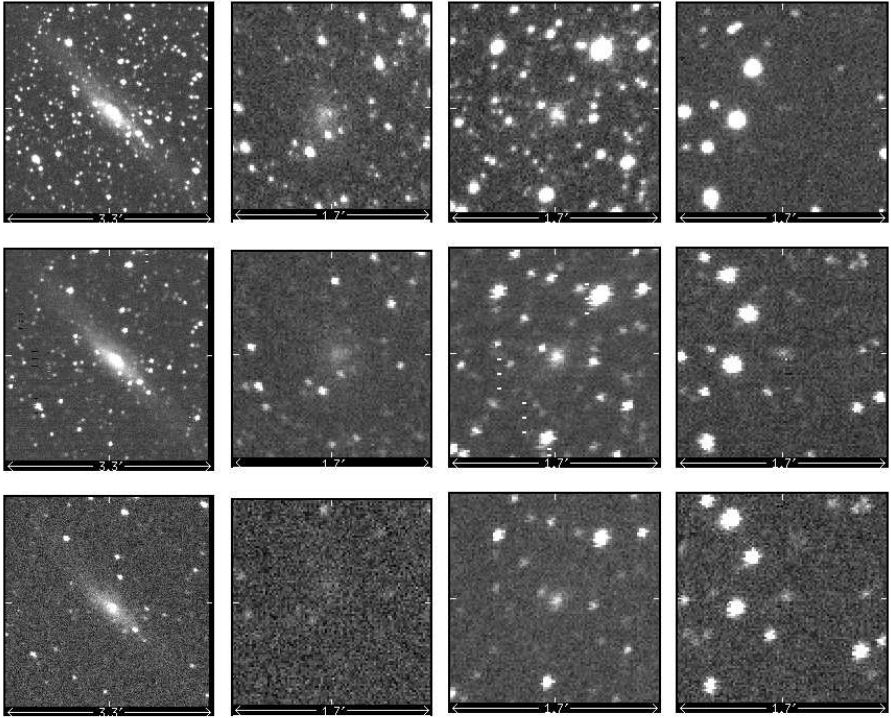}}
\vspace{-.5cm}
\caption{DENIS survey images (before bad pixel filtering)
of four galaxies detected 
blindly in  \hi\ at $|b| \le 5\deg$; the \II\  band image
is at the {\it top
\/}, \J\ in the {\it middle\/} and \K\ at the {\it bottom\/}.}
\label{hidetplot}
\end{figure}

Figure~\ref{hidetplot} shows four examples of the candidates. 
The first galaxy is a nearby  ($v\!=\!1450 \, \rm km \, s^{-1}$) ESO-Lauberts
galaxy (L223-12) at   
$b=+4\fdg8$ and $A_B = 3\fm2$. It is very impressive 
in all three NIR passbands (note the larger image scale for this galaxy,
\ie $3\farcm3$ instead of $1\farcm7$). 
The second galaxy at $(l,b,A_B) = (306\fdg9,+3\fdg6,3\fm3$) is
slightly more distant ($v\!=\!2350\, \rm km \, s^{-1}$). 
This galaxy has also been 
identified in \B\ and is quite distinct in \II\ and \J .
The third galaxy at $(b,A_B) \simeq (-2\fdg9, 4\fm6)$ had been detected by us
as an OFF-signal at $v\!=\!2900\, \rm km \, s^{-1}$ 
during pointed \HI\ observations
in the ZOA. It has no optical counterpart but can be clearly 
seen in all three NIR passbands.  The last example is an uncertain 
NIR counterpart at $(b,A_B) \simeq (+1\fdg5,7\fm5$) of a galaxy
detected in \HI\ at $v\!=\!1450\, \rm km \, s^{-1}$. 
It is barely visible in the \II\  band.

Although the present data is scarce, NIR counterparts of \HI\ detected,
highly obscured galaxies certainly seem to merit a systematic
exploitation for large-scale structure investigations.

\section{Conclusion }

Our pilot study illustrates the promises of using the NIR surveys for 
extragalactic large-scale studies behind the ZOA as well as for the mapping 
of the Galactic extinction. 

{\sl At intermediate latitudes and extinction} 
($5\deg < |b| < 10\deg$, $1^{\rm m} \la A_B \la 4-5^{\rm m}$)
optical surveys remain superior for identifying
galaxies.  However, the NIR luminosities and colours together with
extinction data from the NIR colours will prove invaluable in 
analysing the optical survey data and their distribution in redshift 
space, and in the final merging of these data with existing sky
surveys.  Despite the high extinction and the star crowding 
at these latitudes, \II , \J\ and \K\ photometry from the survey 
data can be successfully performed at these low latitudes and lead,
for instance, to the preliminary $I_c^o$, $J^o$ and $K_s^o$ galaxy 
luminosity functions in A3627.

{\sl At low latitudes and high extinction} 
($|b| < 5\deg$ and $A_B \ga 4-5^{\rm m}$)
the search for `invisible' obscured galaxies on existing DENIS-images 
implicate that NIR-surveys can trace galaxies down to about $|b| \simeq
1\fdg5$. The \J\ band was found to be optimal for identifying galaxies up to
$A_B \simeq 7^{\rm m}$, although this might change in favour of \K\
with the new cooling system.  NIR surveys can hence further reduce the
width of the ZOA. This is furthermore the only tool that permits the 
mapping of early-type galaxies --- tracers of density peaks --- at 
high extinction.  

The combination of two different surveys, \ie NIR data for highly obscured
spiral galaxies detected in a systematic blind \HI\ survey --- a fair
fraction could indeed be re-identified on DENIS-images --- allows the mapping
of the peculiar velocity field in the ZOA through the NIR \tfr.  This will be
pursued as well at intermediate latitudes ($5\deg < |b| < 10\deg$) with
pointed \HI\ observations of optically identified spiral galaxies. About 300
spiral galaxies have alrady been detected (\ca{HIw} \cy{HIw}).

Whether the systematic identification of ZOA galaxies from the DENIS survey
must be performed by visual examination or whether galaxies can be
successfully extracted using classical algorithms (\ca{gam3} \cy{gam3,gam4}) or
artificial neural 
networks (\ca{bertinarnouts} \cy{bertinarnouts}, Bertin, in these
proceedings) or a combination of both requires further
exploration. 
 
\acknowledgements{We thank Jean Borsenberger for providing bias subtracted,
flat fielded DENIS
images, Emmanuel Bertin for supplying recent updates of his
SExtractor software package, and Eric Copet for providing software to display
Figures 4 and 5.}

\vfill
\end{document}